\documentclass[manuscript=article]{achemso}

\usepackage{upgreek}
\usepackage[version=3]{mhchem} 
\usepackage[T1]{fontenc}       
\usepackage{graphicx}					
\usepackage{epstopdf}
\usepackage{amsmath}
\usepackage{color}

\setkeys{acs}{usetitle = true}

\title{Electronic Properties of Interfaces between N-Heterotriangulene Donors and Strong Tetracyanoquinodimethane Acceptors}

\author{Mohsen Ajdari}
\affiliation{Physikalisch-Chemisches Institut, Universit\"at Heidelberg, Im Neuenheimer Feld 253, 69120 Heidelberg, Germany}
\author{Ronja Pappenberger}
\affiliation{Physikalisch-Chemisches Institut, Universit\"at Heidelberg, Im Neuenheimer Feld 253, 69120 Heidelberg, Germany}
\author{Christian Walla}
\affiliation{Interdisziplin\"{a}res Zentrum f\"{u}r Wissenschaftliches Rechnen, Universit\"at Heidelberg, Im Neuenheimer Feld 205A, 69120 Heidelberg, Germany}
\affiliation{Physikalisch-Chemisches Institut, Universit\"at Heidelberg, Im Neuenheimer Feld 253, 69120 Heidelberg, Germany}
\author{Ina Michalsky}
\affiliation{Organisch-Chemisches Institut, Universit\"at Heidelberg, Im Neuenheimer Feld 270, 69120 Heidelberg, Germany}
\author{Friedrich Maa{\ss}}
\affiliation{Physikalisch-Chemisches Institut, Universit\"at Heidelberg, Im Neuenheimer Feld 253, 69120 Heidelberg, Germany}
\author{Milan Kivala}
\affiliation{Organisch-Chemisches Institut, Universit\"at Heidelberg, Im Neuenheimer Feld 270, 69120 Heidelberg, Germany}
\author{Andreas Dreuw}
\affiliation{Interdisziplin\"{a}res Zentrum f\"{u}r Wissenschaftliches Rechnen, Universit\"at Heidelberg, Im Neuenheimer Feld 205A, 69120 Heidelberg, Germany}
\affiliation{Physikalisch-Chemisches Institut, Universit\"at Heidelberg, Im Neuenheimer Feld 253, 69120 Heidelberg, Germany}
\author{Petra Tegeder}
\affiliation{Physikalisch-Chemisches Institut, Universit\"at Heidelberg, Im Neuenheimer Feld 253, 69120 Heidelberg, Germany}
\email{tegeder@uni-heidelberg.de}
\phone{+49 (0) 6221 548475}

\date{\today}
\begin{document}

\begin{abstract}
N-heterotriangulenes (N-HTAs) represent a class of functional molecules with high potential for optoelectronic materials, for example as electron donating compounds in donor/acceptor (D/A) systems. The capability of two different N-HTAs, N-HTA 550 and N-HTA 557, the latter
containing an additional 7-membered ring, to act as electron donors at interfaces with strong tetracyanoquinodimethane (TCNQ and F4TCNQ) acceptors is studied using high-resolution electron energy loss spectroscopy in combination with state-of-the-art quantum chemical calculations. For TCNQ/N-HTA bilayer systems adsorbed on Au(111) Low-energy (< 2.5 eV) electronic transitions which are attributed to charge transfer (CT) states for all four D/A combinations are identified. Based on substantial quantum chemical calculations a generation of ground state CT complexes is  excluded. Instead, CT in the excited state, in which an electron-stimulated CT from the N-HTAs to TCNQs is the underlying process, is proposed. The energies of the CT states are determined by the values of the ionization potential and electron affinity of the involved donor and acceptor.
\end{abstract}
\maketitle

\section{Introduction}
Charge transfer (CT) in strongly interacting donor/acceptor systems is of fundamental importance for the function of organic (opto)electronic devices \cite{Zhang2017, Vandewal2016, Salzmann2015, Lussem2013}, for example as intermediate step in the separation of an exciton into an electron and hole to create charge carriers or to increase the conductivity via doping with suitable electron accepting or donating species.
The electronic and structural properties of many donor/acceptor combinations have been investigated \cite{Bartesaghi2015, Deibel2010, Vandewal2014, Hu2017, Duva2019, Karpov2017, Salzillo2016, Beyer2019, Goetz2014} and three generally different scenarios of charge transfer have been distinguished \cite{Vandewal2016, Salzmann2016}: CT in the excited state \cite{Deibel2010}, integer CT in the ground state leading to ion-pair formation \cite{Pingel2013, Koech2010}, and the generation of a ground state charge transfer complex (CTC) with a large dipole moment.
The latter case is often described in the molecular orbital picture as a hybridization of the frontier orbitals of donor and acceptor molecules \cite{Mendez2015, Salzmann2012}.
In particular, small band gap CTCs with strongly interacting donor/acceptor molecules are interesting from the viewpoint of fundamental research \cite{Niederhausen2020}, and have already found potential application in near-infrared (NIR) photodetectors \cite{Siegmund2017}.

\begin{figure}[htb]
\centering
\resizebox{0.4\hsize}{!}{\includegraphics{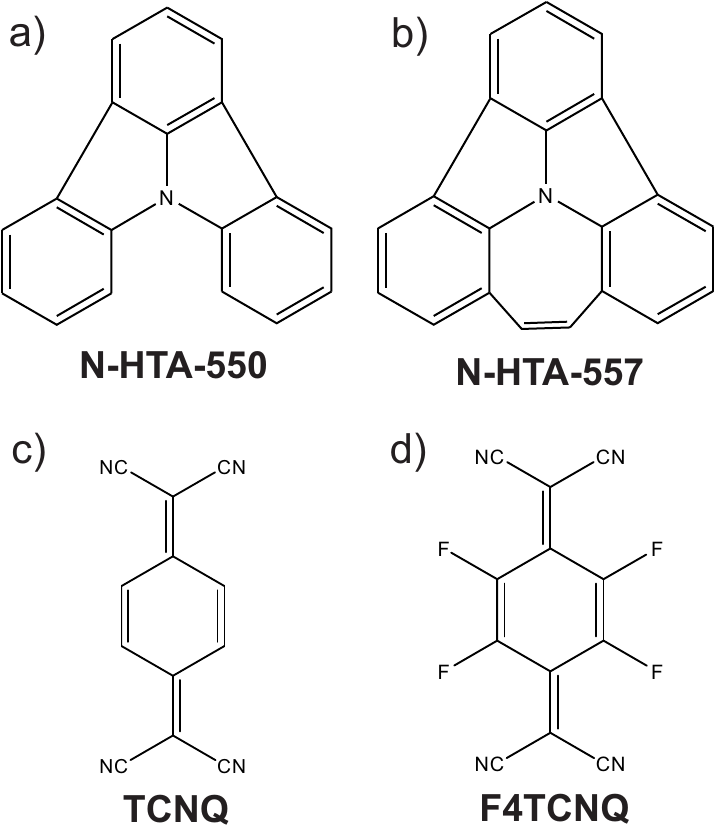}}
\caption{a), b) N-heterotriangulenes (N-HTA) derivatives \cite{Michalsky2022} and the electron acceptors c) TCNQ and d) F4TCNQ  investigated in the present study.}
\label{molecules}
\end{figure}

N-heterotriangulenes (N-HTAs, see Fig. \ref{molecules}), dimethylmethylene-bridged triphenylamines \cite{Hammer2015, Michalsky2022}, represent a class of functional molecules with high potential for application in optoelectronic materials \cite{Meinhardt2016, Hirai2019, Schaub2020, Krug2020}, for example as electron donating compounds in donor/acceptor systems.
Therefore, we recently analyzed the electronic properties and absorption spectra of two N-HTA derivatives, N-HTA-550 and N-HTA-557 (Fig. \ref{molecules}), at the interface to Au(111) and within thin molecular films using vibrational and electronic high resolution electron energy loss spectroscopy (HREELS) in combination with quantum chemical calculations.
All observed spectral features could be assigned to specific electronic states, and moreover, the additional -C=C- bridge forming the 7-membered ring in N-HTA-557 resulted in a pronounced reduction of the optical gap size by 0.9 eV from 3.4 eV in N-HTA-550 to 2.5 eV in N-HTA-557 due to an increase of the $\pi$-conjugated electron system \cite{Ajdari2021b}.

In the present study we again utilize HREELS and quantum chemistry to elucidate possible CT processes between these N-HTA derivatives and the well-known electron acceptors 7,7,8,8-tetracyanoquinodimethane (TCNQ) and 2,3,5,6-tetrafluoro-7,7,8,8-tetracyanoquino- dimethane (F4TCNQ) (see Fig. \ref{molecules}).
We aim at elucidating  the potential of N-HTA as electron donors in general and at understanding the involved CT mechanism.
For bilayer systems of the N-HTA and TCNQ derivatives adsorbed on Au(111) we identified electronic transitions in the low-energy regime (< 2.5 eV) which are attributed to excited CT-states for all donor/acceptor combinations.
In fact, the underlying CT formation process occurs not in the ground, but in the excited state.
Thereby electron-stimulated excitation leads to a CT from the N-HTA donors to the TCNQ acceptors.
The excitation energies of the CT states can be related to the ionization potentials (IPs) and electron affinities (EAs) of the involved compounds.
Thus, the combination F4TCNQ/N-HTA-557 possesses the smallest CT state energy, because the EA of F4TCNQ is higher compared to TCNQ and the IP is smaller for N-HTA-557 than for N-HTA-550.

\section{Experimental and Computational Details}
The HREELS experiments were performed under ultrahigh vacuum conditions at a sample temperature of around 90 K.
Au(111) single crystals were prepared by standard procedure of Ar$^{+}$ sputtering and annealing.
The N-HTAs, TCNQ, and F4TCNQ were deposited from an effusion cell onto the Au(111) sample held at 200 K (for the evaporation temperatures see supporting information).
The coverage was determined by temperature-programmed desorption measurements (see supporting information). HREELS measurements were performed with an incident electron energy of 15 eV. For details see Refs. \cite{ibach1982, Tegeder2012,maass2015, Maass2016, Maass2017}.

All our quantum chemical calculations were performed using the software package Q-Chem 5.3 \cite{qchem}.
The equilibrium geometries of the neutrals and cations of N-HTA-550, N-HTA-557 and the neutrals and anions of TCNQ and F4TCNQ have been optimized at the level of standard density functional theory (DFT) using the long-range corrected CAM-B3LYP exchange-correlation (xc) functional and the cc-pVDZ basis set \cite{yanai2004}.
Similarly, the equilibrium structures of the complexes of the N-HTAs with F4TCNQ and TCNQ have been optimized.
A long-range corrected xc-functional is mandatory for the treatment of charge-transfer states and their excited states \cite{Dreuw2005}.
Therefore, excited states of all neutral and ionic compounds have subsequently been computed at their corresponding equilibrium geometries using linear-response time-dependent DFT (TDDFT) \cite{Dreuw2005} again with the CAM-B3LYP/cc-pVDZ xc-functional and basis set combination.
Ionization potentials and electron affinities have been computed as $\Delta$DFT values, i.e. as difference of the total energies of the neutral and charged species.
Vertical values are computed at the equilibrium geometry of the neutral species also for the cations and anions, while the equilibrium geometry of the charged species is used for the calculation of the their energies to obtain adiabatic IP and EA values.
To model the influence of the environment in the HREELS experiments, the application of a polarizable continuum model (PCM) with a dielectric constant $\epsilon$ of 10 had turned out to be useful \cite{Ajdari2021b} and has been used also in this study unless mentioned otherwise.

\section{Results and Discussion}
Before presenting and discussing the results of the mixed N-HTA donor and TCNQ acceptor systems, we briefly introduce the electronic properties of the bare N-HTA-550 and N-HTA-557 donors and the TCNQ and F4TCNQ acceptor compounds analyzed with HREELS.

\subsection{Electronic transitions of the donor and acceptor compounds}
\begin{figure}[htb]
\centering
\resizebox{0.45\hsize}{!}{\includegraphics{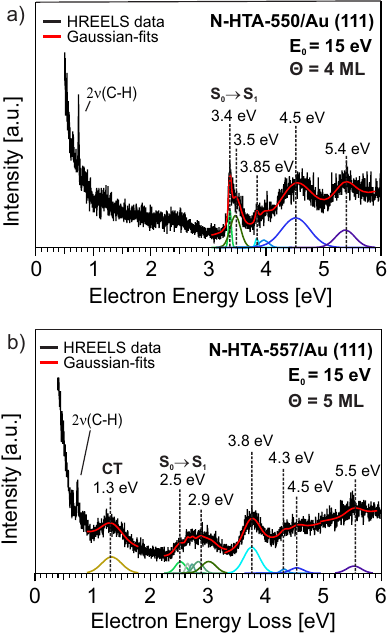}}
\caption{Electronic HREEL spectra of a)  4 ML N-HTA-550/Au(111) and b) 5 ML N-HTA-557/Au(111) measured with an incident electron energy of 15 eV. CT indicates an charge transfer from the N-HTA-557 to the metal surface, creating a positive ion. The electronic transitions were fitted using Gaussian functions (red curves).}
\label{donors}
\end{figure}

Figure \ref{donors} displays the electronic HREELS spectra of the N-HTA derivatives recorded with an incident electron energy of 15 eV. In both cases a multilayer coverage (4 monolayers (ML) N-HTA-550/Au(111) and 5 ML N-HTA-557/Au(111)) has been deposited onto the Au(111) surface to minimize the influence of the metallic substrate on the molecular electronic states.
As previously demonstrated \cite{Ajdari2021b}, for both compounds several peaks are observed, which are assigned to the $S_{0}\rightarrow S_{1}$  transitions (optical gaps) at 3.4 (N-HTA-550) and 2.5 eV (N-HTA-557), respectively as well as higher lying singlet states.  The peak at 1.3 eV in N-HTA-557 has been attributed to a CT-state due to an electron transfer from the N-HTA-557 to the Au(111) substrate.
In the HREELS data of 1 ML N-HTA-557/Au(111) this peak is even more pronounced (data not shown here) caused by the surface sensitivity of HREELS.
Notably, N-HTA-557 is a stronger electron donor compared to N-HTA-550, which is seen in the computed ionization potentials of these compounds (see below), which is 0.4-0.5 eV lower for the first compared to the latter, independent of the computational model and the polarity of the applied solvation model.

This assignment has been supported by the calculated excitation energies at TDDFT/CAM-B3LYP level \cite{Ajdari2021b}.
Because explicit environmental and vibrational effects have been neglected and the applied theoretical model has an intrinsic statistical mean absolute error of 0.3 eV \cite{Shao2019} overestimating excitation energies, one can not expect a quantitative agreement.
Indeed the calculated excitation energies of the neutral compounds deviate from the experimentally determined values consistently by 0.5-0.7 eV, however the qualitative trends and the order of states have been correctly reproduced.

\begin{figure}[htb]
\centering
\resizebox{0.45\hsize}{!}{\includegraphics{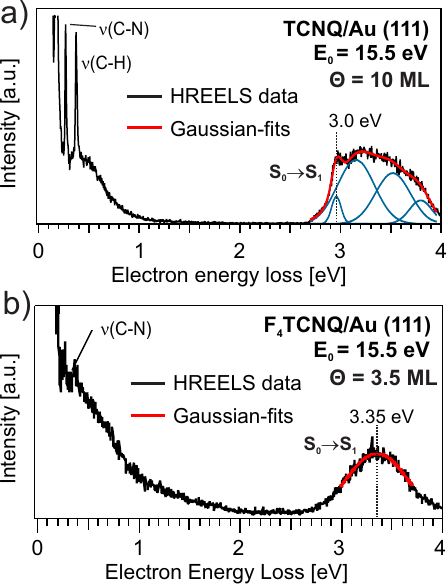}}
\caption{Electronic HREEL spectra of a)  10 ML TCNQ/Au(111) and b) 9 ML F4TCNQ/Au(111) measured with an incident electron energy of 15 eV. The electronic transitions were fitted using Gaussian functions (red curves).}
\label{acceptors}
\end{figure}

The corresponding HREEL data for the acceptor molecules TCNQ and F4TCNQ are shown in Fig. \ref{acceptors}.
For TCNQ a pronounced peak including vibronic contributions at 3.0 eV is associated with the optical gap in agreement with UV/vis data of TCNQ in solution (3.1 eV \cite{Ma2016}).
F4TCNQ shows a dominant peak at 3.35 eV, which is assigned to the $S_{0}\rightarrow S_{1}$ transition in accordance with literature \cite{Ma2016}.
The assignment is further substantiated by our computed excitation energies at TDDFT/CAM-B3LYP level, which are again consistently overestimated by about 0.5 eV (Table \ref{tab:ion_spec}).
The S$_1$ state of TCNQ is the typical $\pi\pi^*$ state and has an excitation energy of 3.70 eV and a large oscillator strength.
In contrast, the S$_1$ state of F4TCNQ is optically forbidden with an excitation energy of 3.2 eV, while the S$_2$ state is the typical $\pi\pi^*$ state with an excitation energy of 3.63 eV and large oscillator strength which corresponds thus to the visible peak the HREELS spectrum.
Due to F-substitution in TCNQ the electron affinity (EA) of F4TCNQ is much higher in comparison to TCNQ (EA(TCNQ) = 4.32 eV \cite{Kanai2009};  EA(F4TCNQ) = 5.1--5.4 eV \cite{Yoshida2015, Kanai2009, Gao2001}).

\subsection{Electronic transitions of the N-HTA/TCNQ donor/acceptor systems}
\begin{figure}[htb]
\centering
\resizebox{0.45\hsize}{!}{\includegraphics{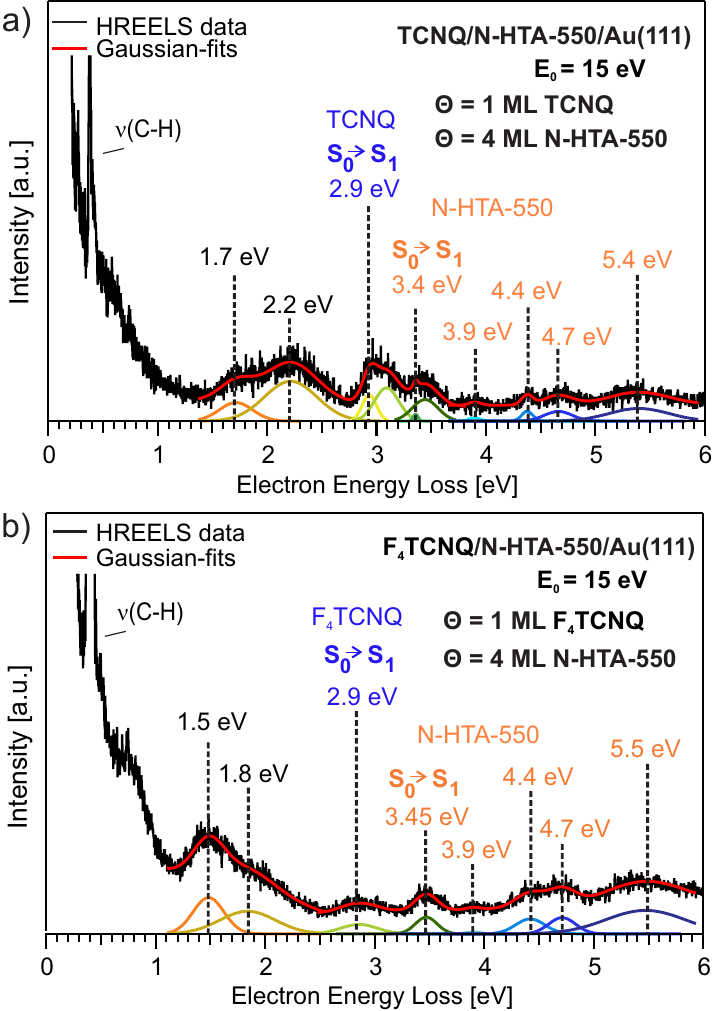}}
\caption{Electronic HREEL spectra of (a) 1 ML TCNQ deposited on 4 ML N-HTA-550 on Au(111) and (b) 1 ML TCNQ deposited on 4 ML N-HTA-557 on Au(111). The peaks were fitted with Gaussian functions (red) and the positions
are given in eV. The transitions associated with TCNQ or F4TCNQ and N-HTA-550
are marked in blue and orange, respectively.}
\label{N-HTA550_acceptor}
\end{figure}

 To study possible CT between the N-HTAs and the acceptors we prepared a multilayer of the donor N-HTA on the Au(111) followed by deposition of approximately 1 ML of the acceptor compound, since it is well-known that at least for F4TCNQ a charge transfer from the Au(111) substrate to F4TCNQ is occurring when adsorbing it directly on Au(111) and thus leading to the formation of negatively charged F4TCNQ \cite{Koch2005, Rangger2009, Faraggi2012, Gerbert2018}. Figure \ref{N-HTA550_acceptor} displays the HREEL data obtained from N-HTA550 with both acceptors.
For N-HTA-550/TCNQ as well as for N-HTA-550/F4TCNQ system low-energy transitions are observed which are not found in the single components. For both systems a double peak structure at 1.7 and 2.2 eV for N-HTA-550/TCNQ and at 1.5 and 1.8 eV for N-HTA-550/F4TCNQ is detected. Notably, for  the stronger electron acceptor F4TCNQ the transitions are located at lower energies.
\begin{figure}[htb]
\centering
\resizebox{0.45\hsize}{!}{\includegraphics{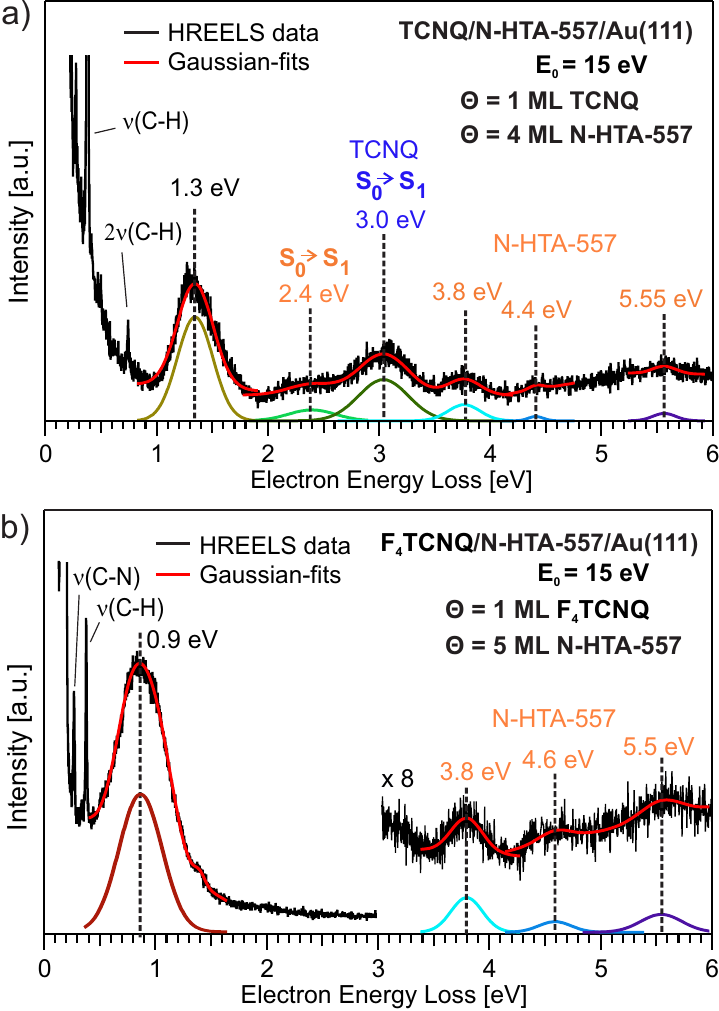}}
\caption{Electronic HREEL spectra of (a) 1 ML TCNQ deposited on 4 ML N-HTA-557 on Au(111) and (b) 1 ML TCNQ deposited on 5 ML N-HTA-557 on Au(111). The peaks were fitted with Gaussian functions (red) and the positions
are given in eV. The transitions associated with TCNQ or F4TCNQ and N-HTA-557
are marked in blue and orange, respectively.}
\label{N-HTA557_acceptor}
\end{figure}

The corresponding spectra for N-HTA557 with both acceptors are displayed in Fig. \ref{N-HTA557_acceptor}.
Also for the N-HTA557/acceptor systems low-energy contributions are seen, apart from the transitions associated with the single components.
In the case of N-HTA-557/TCNQ a much more pronounced peak at 1.3 eV compared to the bare N-HTA-557/Au(111) is observed.
For N-HTA-557/F4TCNQ an even more distinct peak at 0.9 eV is visible.
As found for the N-HTA-550/F4TCNQ system, the transition in N-HTA-557/F4TCNQ is at a lower energy.
The energetic positions and strengths of the lowest peaks seem to be diretly related to the donor and acceptor strengths of the individual compounds of the complexes.
Therefore, these low-energy transitions most likely correspond to CT states at the donor/acceptor interface.
To further corroborate this hypothesis we performed state-of-the-art DFT calculations on the electronic properties and spectra of the possibly involved molecular species.

\section{Computational Results}

As a first step of the computational investigation, let us establish the electronic structure of the ground state of the complexes of the N-HTA and TCNQ derivatives.
Based on chemical intuition, it is expected that N-HTA-557 is a stronger electron donor than N-HTA-550 and F4TCNQ is a stronger electron acceptor than TCNQ.
This is corroborated by our computations of the corresponding vertical ionization potentials (IP) of N-HTA-557 and N-HTA-550, which amount to 7.38 and 7.81 eV in the gas phase respectively.
To estimate whether nuclear dynamics and accompanying changes of the molecular geometry have a large influence on the IPs, we also computed the adiabatic IPs at the optimized equilibrium geometries of the cations.
At the level of $\Delta$DFT/CAM-B3LYP/cc-pVDZ they possess values of 7.26 and 7.76 eV for N-HTA-557 and N-HTA-550, respectively (Table \ref{tab:IPs_EAs}).
Since they are only negligibly smaller than the vertical values, nuclear motion has practically no influence on the IPs.
The same trend is seen for the electron affinities (EA) of F4TCNQ and TCNQ.
Their vertical EAs amount to 3.52 and 3.01 eV and their adiabatic ones to 3.72 and 3.21 eV in the gas phase (Table \ref{tab:IPs_EAs}).
Also here, nuclear motion leads to only marginal stabilization of the anions.

Including solvation by using a polarizable continuum model with increasing dielectric constants shows the expected stabilization of the charged species, resulting in a decrease of the IPs and an increase of the EAs (Table \ref{tab:IPs_EAs}).
Using a dielectric constant $\epsilon$ of 5 reveals a strong drop of more than 1 eV of the IP of N-HTA-557 from 7.38 eV to 6.11 eV, while further increase of $\epsilon$ up to 80 lowers the IP only slightly to 5.81 eV.
The same behaviour is observed for the EA of F4TCNQ, a large increase from 3.52 to 4.46 eV for $\epsilon$=5 and then only a small further growth to 4.66 for $\epsilon$=80.
Since the correct value of the dielectric constant $\epsilon$ for modelling the environment in the HREELS experiments is not known, it is important to note that the most sensitive quantities are largely independent of its actual value.

\begin{table}[]
    \centering
    \begin{tabular}{l|cccccc}\hline\hline
    & gas phase & $\epsilon$=5 & $\epsilon$=10 & $\epsilon$=20 & $\epsilon$=40 & $\epsilon$=80 \\\hline

    \multicolumn{7}{l}{\em Ionization Potentials} \\
        NHTA550 & 7.81 &  &  & & 6.23 &\\
                &(7.76)&  &  & & (6.18)& \\\hline
        NHTA557 & 7.38 & 6.11 & 5.95 & 5.87 & 5.83 & 5.81\\
                &(7.26)&(5.98)&(5.82)&(5.74)&(5.71)&(5.69)\\\hline
    \multicolumn{7}{l}{\em Electron Affinities} \\
        TCNQ & 3.01 &  &  & & 4.21 & \\
             &(3.21)&  &  & & (4.42)&\\
        F4TCNQ & 3.52 & 4.46 & 4.57 & 4.62 & 4.65 & 4.66 \\
               &(3.72)&(4.66)&(4.77)&(4.82)&(4.85)&(4.86)\\\hline\hline

    \end{tabular}
    \caption{Vertical and adiabatic (in parenthesis) ionization potentials of N-HTA-550 and N-HTA-557 as well as the electron affinities of TCNQ and F4TCNQ given in eV calculated at the level of $\Delta$DFT/CAM-B3LYP/cc-pVDZ using a polarizable continuum model for solvation with increasing dielectric constants up to $\epsilon$=80 modelling polar solvants like water.}
    \label{tab:IPs_EAs}
\end{table}

When a complex of N-HTA-557 and F4TCNQ is formed, one may expect the formation of a charge-transfer complex and transfer of an electron from N-HTA-557 to F4TCNQ forming a cation and an anion already in the electronic ground state.
The corresponding required energy is often estimated via a point-charge model as E$_{CT}$=IP-EA-1/R.
Using this formula and the computed values for the IP and the EA for $\epsilon$=10, for example, the intermolecular distance for which E$_{CT}$ becomes negative, i.e. when electron transfer will occur, needs to be smaller than 11.7 \AA.
In the optimized equilibrium geometry of the N-HTA-557$\cdot$F4TCNQ complex the intermolecular distance is much smaller with only 3.2 \AA{} in an almost coplanar arrangement.
In contrast to expectation, the computed dipole moment of the complex is however very small with only 2.52 D using a PCM with $\epsilon$=10 and the Mulliken charge distribution clearly indicates that no charge transfer takes place between N-HTA-557 and F4TCNQ in the electronic ground state.
One important factor neglected in the point-charge estimate is the energy needed to desolvate the free ions to form the complex.
In general, this discrepancy between the point-charge model and the actual finding, also demonstrates its inappropriateness for such large molecules at short intermolecular distances.
Even when the dielectric constant is increased up to $\epsilon$=80, no charge transfer complex is observed.
Therefore, the formation of ground-state charge-transfer complex can be excluded to be relevant for the interpretation of the HREELS experiments.

Nevertheless, due to the experimental setup and the presence of the gold surface, ion formation via electron transfer to or from the surface can generally not be excluded.
To investigate whether the observed low-energy peaks in the HREELS spectra can originate from N-HTA-550 and N-HTA-557 cations or TCNQ and F4TCNQ anions, we computed the electronic excited states of the neutral and ionic individuals (Table \ref{tab:ion_spec}).
While the spectra of the neutral individual species agree qualitatively very well with the measured ones (Figs. \ref{donors} and \ref{acceptors}), and which have been assigned already previously \cite{Michalsky2022}, the excited states of the anions and cations could not be identified in the experimental spectra of neither the individual compounds nor the complexes.
Therefore, also the presence of cationic N-HTA-550 and N-HTA-557 as well as anionic F4TCNQ and TCNQ species can be excluded based on our computational results.

\begin{table}[]
    \centering
    \begin{tabular}{l|cccc}\hline\hline
    & N-HTA-550 & N-HTA-557 & TCNQ & F4TCNQ \\\hline

    \multicolumn{5}{l}{\em Neutrals} \\
        S$_1$ & 4.07 (0.11) & 2.99 (0.03) & 3.70 (1.86) & 3.20 (0.00)\\
        S$_2$ & 4.45 (0.08) & 3.60 (0.06) & 3.98 (0.00) & 3.63 (2.08)\\
        S$_3$ & 4.86 (0.19) & 4.38 (0.17) & 4.99 (0.00) & 4.77 (0.00)\\
        S$_4$ & 5.01 (0.09) & 4.49 (0.05) & 5.16 (0.00) & 4.77 (0.00)\\
        S$_5$ & 5.29 (0.37) & 4.65 (0.15) & 5.26 (0.00) & 4.93 (0.00)\\
        T$_1$ & 2.90 & 1.81 & 1.28 & 1.15 \\
        T$_2$ & 3.17 & 2.93 & 3.09 & 2.50 \\\hline
    \multicolumn{5}{l}{\em Cations/Anions} \\
        D$_1$ &  0.83 (0.01) & 1.51 (0.00) & 1.77 (0.36) & 1.71 (0.37) \\
        D$_2$ &  0.94 (0.00) & 1.65 (0.00) & 3.29 (0.00) & 2.87 (0.00) \\
        D$_3$ &  1.83 (0.19) & 2.04 (0.06) & 3.54 (0.01) & 3.63 (0.52)\\
        D$_4$ &  2.33 (0.04) & 2.15 (0.02) & 3.62 (0.47) & 3.97 (0.00)\\
        D$_5$ &  3.00 (0.01) & 2.51 (0.10) & 3.88 (0.00) & 4.08 (0.01) \\\hline\hline

    \end{tabular}
    \caption{Excitation energies of the neutral and cationic species of N-HTA-550 and N-HTA-557 at TDDFT/CAM-B3LYP/cc-pVDZ level as well as of the neutral and anionic species of TCNQ and F4TCNQ.}
    \label{tab:ion_spec}
\end{table}

\begin{figure}[htb]
\centering
\resizebox{0.45\hsize}{!}{\includegraphics{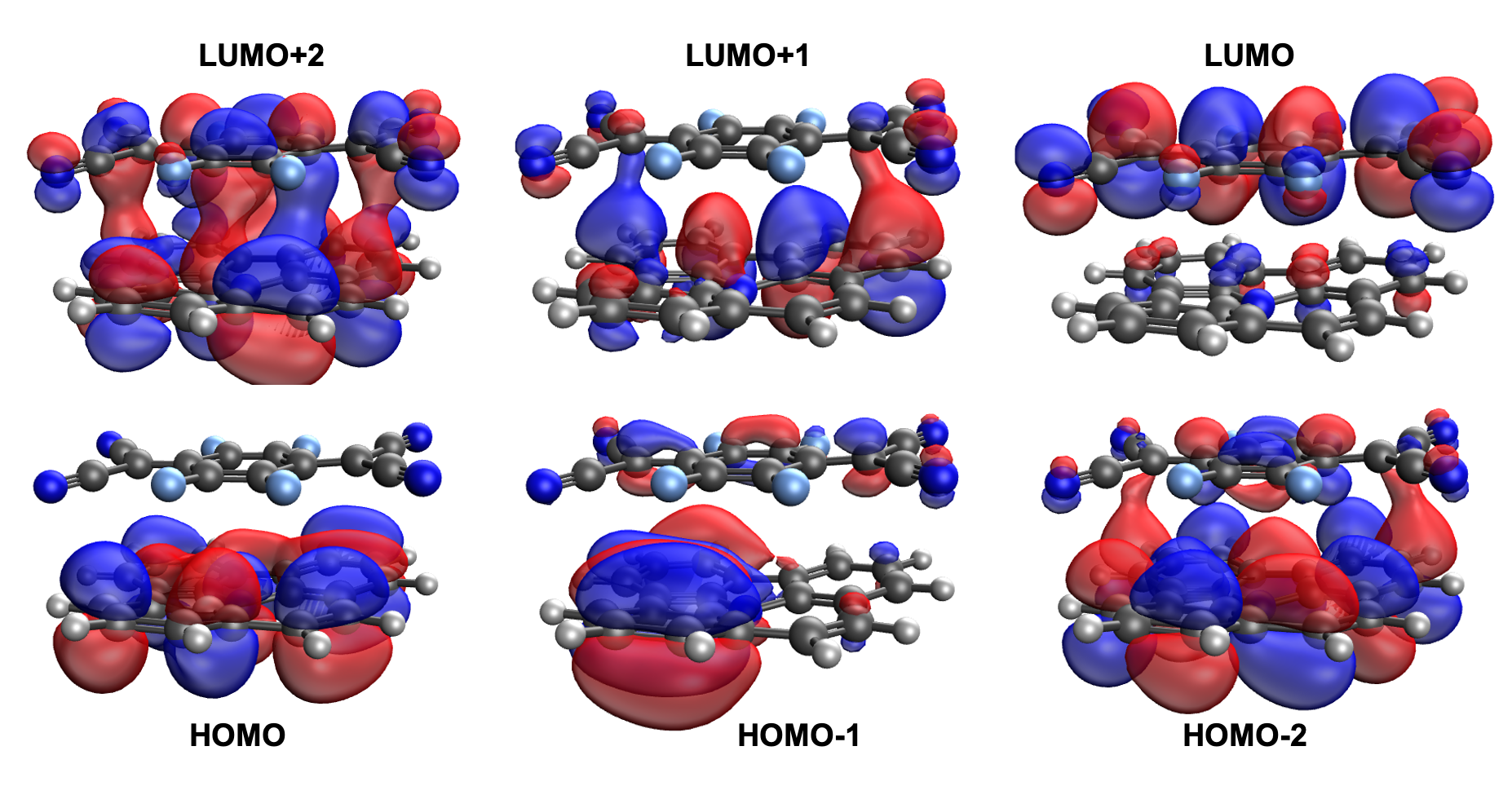}}
\caption{Frontier molecular orbitals of the N-HTA-557$\cdot$F4TCNQ complex at the level of DFT/CAM-B3LYP.}
\label{orbitals}
\end{figure}

Turning to the computation of the vertical excited states of the donor/acceptor complexes of the N-HTA derivatives and both TCNQ and F4TCNQ, the N-HTA-550$\cdot$TCNQ complex has the donor compound with the higher IP and the acceptor compound with the lower EA.
Therefore, electron stimulated charge transfer is expected to occur at higher excitation energies than in the other complexes.
The lowest excited S$_1$ state of N-HTA-550$\cdot$TCNQ has an excitation energy of 2.12 eV and corresponds in the molecular orbital picture to the excitation of a single electron from the highest occupied (HOMO) to the lowest unoccupied molecular orbital (LUMO) (Tab. \ref{tab:ExE_Complexes}).
Since the HOMO is mostly located at the N-HTA-550 molecule and the LUMO at the TCNQ, this state corresponds to a charge-transfer excited state.
The HOMO and LUMO of all for investigated complexes are practically indistinguishable from those shown in Fig. \ref{orbitals} for the N-HTA-557$\cdot$F4TCNQ complex.
The second excited state of N-HTA-550$\cdot$TCNQ exhibits a computed excitation energy of 2.67 eV at the level of TDDFT/CAM-B3LYP and corresponds essentially to the transition of an electron from the HOMO-1 to the LUMO, and corresponds thus also to a charge-transfer excited state.
Similar to the isolated molecules, the applied methods can be expected to overestimate the excitation energies of these states as well, and thus the peaks seen in the HREELS spectrum at 1.7 and 2.2 eV can most likely be assigned to these two charge-transfer excited states.
Although the absolute values of the excitation energies are overestimated by about 0.5 eV, the gap between these states with 0.5 eV is very well reproduced.

Turning to the N-HTA-550$\cdot$F4TCNQ complex, F4TCNQ has a higher electron affinity than TCNQ by about 0.5 to 0.3 eV depending on the polarity of the environment (Table \ref{tab:IPs_EAs}), and thus, the excited CT states can be expected to lie energetically lower by approximately this amount of energy.
Indeed, the excitation energies obtained at the theoretical level of TDDFT/CAM-B3LYP for the S$_1$ and S$_2$ states are 1.70 and 2.18 eV, almost precisely 0.5 eV lower than those of N-HTA-550$\cdot$TCNQ.
The electronic transitions possess the same charcter, and in the molecular orbital picture the same molecular orbitals are involved.
In the HREELS spectrum of N-HTA-550$\cdot$F4TCNQ two peaks are observed at 1.5 and 1.8 eV, i.e. 0.2-0.4 eV lower than in the spectrum of N-HTA-550$\cdot$TCNQ, which is consistent with our assignment to these two low-lying excited CT states.

In the N-HTA-557$\cdot$TCNQ complex, N-HTA-557 possesses a IP that is 0.4 eV lower than the one of N-HTA-550, and hence the excited CT states should again be lower than those of the N-HTA-550$\cdot$TCNQ complex.
Indeed this is the case for the first excited S$_1$ state with an excitation energy of 1.64 eV.
Again, this state corresponds to the transition of an electron from the HOMO at N-HTA-557 to the LUMO located at the TCNQ moiety.
The second excited S$_2$ state, however, is found at higher excitation energy of 2.71 eV, and does not correspond to a pure charge-transfer excited state as it contains admixtures of locally excited determinants.
This is in agreement with the observed HREELS spectrum of the N-HTA-557$\cdot$TCNQ complex, which contains only a single peak in the low-energy region at 1.3 eV, and which can thus be assigned to originate from the S$_1$ CT state.

Since N-HTA-557 and F4TCNQ possess the lowest IP and highest EA of the four species studied, the N-HTA-557$\cdot$F4TCNQ is expected to possess the lowest lying CT excited states of all studied complexes.
According to our TDDFT/CAM-B3LYP calculations, the S$_1$ state corresponds to the HOMO to LUMO transition (Fig. \ref{orbitals}) and possesses an excitation energy of only 1.23 eV.
Again, the second excited state lies energetically quite high at 2.31 eV and can be understood as a HOMO-1 to LUMO transition, thus also corresponding to a CT excited state.
In the experimental HREELS spectrum of N-HTA-557$\cdot$F4TCNQ, one intense peak is observed at 0.9 eV, which clearly corresponds to its S$_1$ CT state.

\begin{table}[]
    \centering
    \begin{tabular}{l|cccc}\hline\hline
    & N-HTA-550$\cdot$TCNQ & N-HTA-550$\cdot$F4TCNQ & N-HTA-557$\cdot$TCNQ & N-HTA-557$\cdot$F4TCNQ \\\hline
S$_1$ &  2.12 (0.01) &  1.70 (0.00) & 1.64 (0.01) & 1.23 (0.00) \\
        S$_2$ & 2.67 (0.05) & 2.18 (0.02) & 2.71 (0.08) & 2.31 (0.01)\\
        S$_3$ & 2.74 (0.17) &  2.37 (0.16) & 2.77 (0.17) & 2.49 (0.18) \\
        S$_4$ & 3.28 (0.97) & 3.08 (0.50) & 3.07 (0.44) & 2.77 (0.26)\\
        S$_5$ & 3.63 (0.02) & 3.17 (0.54) & 3.19 (0.10) & 3.16 (0.19) \\
S$_6$ & 3.99 (0.00) & 3.43 (0.01) & 3.61 (0.22) & 3.22 (0.16)\\
S$_7$ & 4.10 (0.19) & 3.80 (0.14) & 3.66 (0.25) & 3.41 (0.41)\\
S$_8$ & 4.26 (0.07) & 4.06 (0.17) & 3.77 (0.12) & 3.57 (0.14)\\
S$_9$ & 4.49 (0.17) & 4.46 (0.22) & 4.09 (0.01) & 3.70 (0.12)\\
S$_{10}$ & 4.85 (0.26) & 4.47 (0.08) & 4.46 (0.29) & 4.39 (0.08)\\\hline
        T$_1$ & 1.43 & 1.26 & 1.40 & 1.18\\
        T$_2$ & 2.09 & 1.65 & 1.59 & 1.29\\\hline\hline

    \end{tabular}
    \caption{Excitation energies of the equilibrium geometries of the four investigated complexes computed at the theoretical level of TDDFT/CAM-B3LYP/cc-pVDZ using a polarizable continuum model with a dielectric constant of  $\epsilon$=10.}
    \label{tab:ExE_Complexes}
\end{table}

\section{Conclusion}
Two N-heterotriangulene (N-HTA) derivatives, the N-HTA-550 and N-HTA-557, which act as electron donors (D) in combination with the strong tetracyanoquinodimethane (TCNQ and F4TCNQ) acceptors (A) adsorbed in bilayers on Au(111) have been
investigated using HREELS, focusing on charge transfer (CT) at the TCNQ/N-HTA interface. For all four D/A-combinations low-energy (< 2.5 eV) electronic transitions have been detected.
Our extensive quantum chemical calculations allowed us to assign the spectra of four complexes formed by N-HTA-550 and N-HTA-557 with TCNQ and F4TCNQ.
We could exclude the presence of CT complexes in the electronic ground state, since even when modelling  highly polar environments electron transfer from the N-HTA donors to the TCNQ/F4TCNQ acceptors could not be observed.
Also, characteristic spectroscopic features of the charged species could not be observed in the experiments further corroborating this finding.
Instead, the newly occurring spectral features in the low-energy region of the HREELS spectra upon complex formation could be unambiguously determined to originate from electron-stimulated charge transfer from the N-HTA derivatives to TCNQ or F4TCNQ.
Although the computed absolute values of the excitation energies deviate from the experimentally determined ones by typical 0.3-0.5 eV, their relative energies are nicely consistent with the computed values of the ionization potentials and electron affinities of the individual compounds, which are the leading quantities determining the energies of the CT states.

\section{Acknowledgments}
Funding by the German Research Foundation through
the collaborative research center SFB 1249 "N-Heteropolycycles as Functional Materials" (project number 281029004-SFB 1249, projects A05, B01, and B06) is gratefully acknowledged.
The authors acknowledge support by the state of Baden-W\"{u}rttemberg through bwHPC
and the German Research Foundation (DFG) through grant no INST 40/575-1 FUGG (JUSTUS 2 cluster).

\section{Conflict of Interest}
There are no conflicts to declare.

\bibliography{N-HTA_lit}{}


\end{document}